# Spin Diffusion and Relaxation in Solid State Spin Quantum Computer

G.P. Berman, B.M. Chernobrod, V.N. Gorshkov,

Theoretical Division, Los Alamos National Laboratory,

Los Alamos, New Mexico 87545

V.I. Tsifrinovich

IDS Department, Polytechnic University, Brooklyn, NY 11201

**Abstract**

The processes of spin diffusion and relaxation are studied theoretically and numerically for quantum computation applications. Two possible realizations of a spin quantum computer (SQC) are analyzed: (i) a boundary spin chain in a 2D spin array and (ii) an isolated spin chain. In both cases, spin diffusion and relaxation are caused by a fast relaxing spin located outside the SQC. We have shown that in both cases the relaxation can be suppressed by an external non-uniform magnetic field. In the second case, our computer simulations have revealed various types of relaxation processes including the excitation of a random distribution of magnetic moments and the formation of stationary and moving domain walls. The region of optimal parameters for suppression of rapid spin relaxation is discussed.

1. **Introduction**

A solid-state quantum computer, operating with electron spins as qubits, is one of the most popular proposals in the theory of quantum computation. (See, for example, [1-7]). For a practical implementation of quantum computation in a spin quantum computer (SQC) the relaxation rate in the spin system must be as small as possible. Spin diffusion is, commonly, one of the most important factors in the electron relaxation processes. (See, for example, [8-10]). Experiments [10] have demonstrated an effective suppression of spin diffusion and relaxation using non-uniform magnetic fields. In our previous work [11] we developed a quasi-classical computational model that successfully describes experiments [10]. In this paper, we apply our model to two possible implementations of a SQC. In section 2 we describe the equation of motion we use. In section 3 we assume that a boundary spin chain in a 2D-spin array is used as a 1D-SQC. While this SQC is free from fast relaxing spins (FRS), FRS can appear at some distance from QSC. Due to the dipole-dipole interaction between the spins, the process of spin diffusion may greatly increase the spin relaxation rate in a SQC. In section 4 we consider an isolated 1D spin chain. We simulate the process of spin relaxation in a SQC and find how to suppress relaxation using a non-uniform magnetic field. Our results can be applied to many SQCs, including an SQC based on self-assembled monolayers [12-14].

## 2. Equations of motion

We use the same dimensionless equations of motion in the rotating frame as in [11]:

$$\frac{d\vec{m}_p}{d\tau} = [\vec{\Omega}_p \times \vec{m}_p],$$

$$\vec{\Omega}_p = \left( -\frac{\chi}{2} \sum_{k \neq p} A_{kp} m_{kx} + \xi'_p m_{py},\ -\frac{\chi}{2} \sum_{k \neq p} A_{kp} m_{ky} - \xi'_p m_{px},\ \delta_p + \chi \sum_{k \neq p} A_{kp} m_{kz} \right), \quad (1)$$

where

$$\vec{m}_p = \vec{\mu}_p / \mu_B,\quad \dot{\vec{m}}_p = d\vec{m}_p / d\tau,\quad \tau = \xi_0 \omega_0 t,\quad \xi'_p = \xi_p / \xi_0,$$

$$\delta_p = \frac{B_{fpz} - \langle B_{fz} \rangle}{\xi_0 B_0},\quad \chi = \frac{\mu_0}{4\pi} \frac{\mu_B}{\xi_0 a^3 B_0},\quad A_{kp} = \frac{3\cos^2 \theta_{kp} - 1}{(r_{kp}/a)^3}. \quad (2)$$

Here $\vec{\mu}_p$ is the magnetic moment of the spin "$p$", $\omega_0 = \gamma B_0$ is the frequency of the rotating frame, $B_0 = B_{ext} + \langle B_{fp} \rangle$, $\vec{B}_{ext}$ is the uniform external magnetic field which points in the positive $z$-direction, $\vec{B}_f$ is the non-uniform magnetic field (which is produced, for example, by ferromagnetic particle), $\vec{B}_{fp}$ is the non-uniform magnetic field acting on spin "$p$", $\langle ... \rangle$ represents an average over the spin system, $\xi_0 \omega_0 = 1 s^{-1}$, $\tau$ is the dimensional time, $\xi'_p$ is the dimensionless relaxation parameter, $a$ is the average distance between spins, $\theta_{kp}$ is the polar angle of the unit vector which points from the spin $k$ to the spin $p$, $\chi$ is the parameter of the dipole-dipole interaction. The computational algorithm, which we use, has been described in detail in [11]. Below we discuss the results of our computer simulations.

### 3. Spin relaxation in 2D- spin systems

We consider the following initial conditions for the relaxation process: at $\tau = 0$ the z-components of $\vec{m}_p$ are given random values between $-0.8$ and $-1$ with a random direction for the transverse components. The spin coordinates in the x-y plane are given by

$$\vec{r}_p = a(l_{p1}\vec{i} + l_{p2}\vec{j}) + \delta\vec{r}_p, \qquad (3)$$

where $-(n-1)/2 \le l_{p1,2} \le (n-1)/2$, $n$ is an odd number (in order to be able to put an FRS in the middle of the 2D plane), the total number of spins is $N = n^2$, $\vec{i}$ and $\vec{j}$ are the unit vectors in the positive x- and y- directions, $\delta x_p$, $\delta y_p$ - are random numbers, which do not exceed $0.05a$, and $\delta\vec{r}_p = \delta\vec{x}_p + \delta\vec{y}_p$.

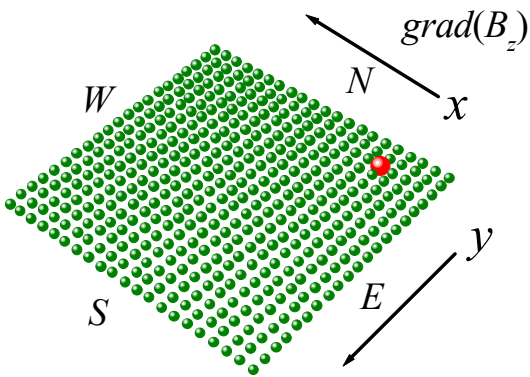 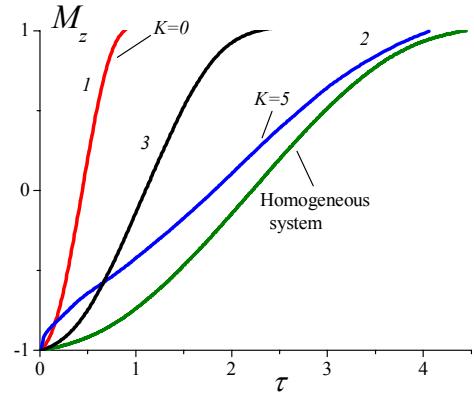

Fig. 1                                                                 Fig. 2

Fig. 1. A 2D spin system with FRS. FRS is shown as a big sphere, other spins are represented by small spheres. SQC is the boundary spin chain, which is parallel to the X-axis.

Fig. 2. Curves 1 and 3 show the dependence $M_z(\tau)$ for $K = 0$: 1- for $\alpha = 1$, $\beta/N = 4$; 3 – for $\alpha = 0$, $\beta/N = 2$. Curve 2 corresponds to $K = 5$, $\alpha = 1$, $\beta/N = 4$. The "homogeneous system" curve corresponds to $K = 0$, $\alpha = \beta = 1$. For all curves $N = 441$ ($n = 21$).

We consider a single FRS with the relaxation parameter $\xi'_p = \beta$. All other slowly relaxing spins (SRS) have the relaxation parameter $\xi'_p = \alpha \ll \beta$. We assume that the magnetic field $B_{fpz}$ increases with the increasing of $x$ and we approximate the value of $\delta_p$ as

$$\delta_p = \frac{aG}{\xi_0 B_0}(x_p/a), \quad G = \left.\frac{\partial B_{fpz}}{\partial x}\right|_{x=0}. \tag{4}$$

Our simulations show that similar to the 3D case [11] the suppression of the relaxation rate depends on the ratio $K = G'/\chi$, where $G' = aG/(\xi_0 B_0)$ is the characteristic frequency difference due to the magnetic field gradient.

Fig. 1 demonstrates one of the studied of 2D- spin systems with a FRS. Our simulations show that at $K = 0$ the relaxation process in the whole 2D- spin system does not depend on the location of the FRS. The relaxation time

$$\tau_1 = \frac{1}{2}\int_0^\infty [1 - M_z(\tau)], \\ M_z(\tau) = \sum_p m_{pz}(\tau) / \sum_p |m_{pz}(0)|\, d\tau \tag{5}$$

is defined by the same expression as applies to the case of a 3D-system [11]:

$$\tau_1 = \frac{2.2}{\alpha + \beta/N}. \tag{6}$$

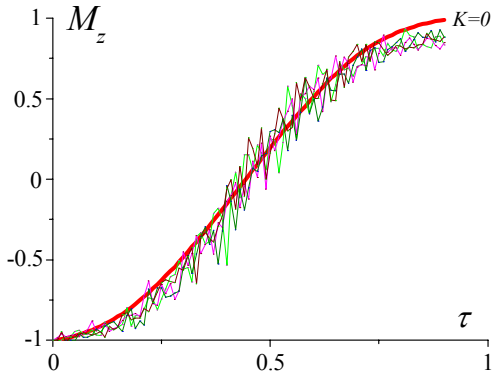

Fig. 3. Dependence $M_z(\tau)$ for $K = 0$. Smooth solid line corresponds to the whole spin system, broken lines correspond to boundary chains $N$, $S$, $W$ and $E$. $\alpha = 1$, $\beta/N = 4$, $N = 441$.

In Fig. 2, curves 1 and 3 demonstrate the dependence $M_z(\tau)$ for $K = 0$. Note that for $K = 0$ the spin relaxation for all boundary spin chains ($N$, $S$, $W$, $E$ in Fig. 1) is approximately the same as relaxation of the whole spin system. (See Fig. 3). We believe that this result is valid for cases in wich the number of spins $N$ does not exceed the threshold value $N_c = n_c^2$, which depends on the effective constant of the dipole-dipole interaction. A rough estimate for $n_c$ can be obtained in the following way. For example, if an FRS is located at the center of a 2D- spin system, the dipole-dipole interaction between the FRS "$q$" and the most remote spin "$k$" must satisfy the condition

$$\Omega_{kz}^{(q)} = \chi A_{kq} \gg 2\pi/\tau_1 \text{ or } 8\chi/n_c^3 \gg \pi/\tau_1 \qquad (7)$$

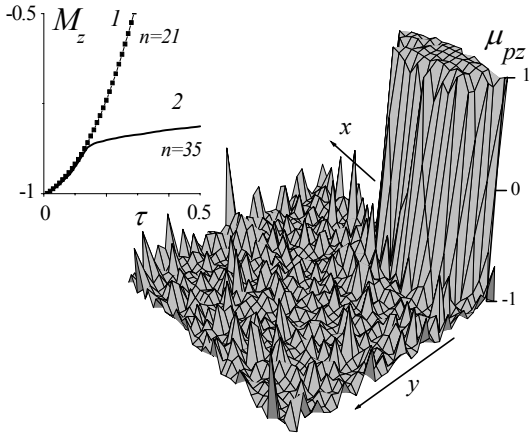

Fig. 4. Formation of a magnetic domain in a large spin system. $\tau = 0.5$, $M_z = -0.865$.

Parameters: $\chi = 1.6 \times 10^5$ ($a = 48 nm$), $\beta/N = 4$, $N = 1225$ ($n = 35$). Curve 2 in insert demonstrates the dependence $M_z(\tau)$, and curve 1 is identical to curve 1 in Fig. 1.

in order for Eq. (6) to be valid. If this condition is not fulfilled, the spin diffusion becomes inhomogeneous over the spin system. The SRS that are located near the FRS form a magnetic domain. (See Fig. 4). At the boundary of this domain, the magnetic field $\vec{B}_{dp}$ has a sufficient gradient to suppress the spin diffusion outside of the domain.

The effect of reducing the spin relaxation rate due to the magnetic field gradient can be described as following. Consider, for example, three spin chains parallel to the $y$-axis (in Fig. 5 they correspond to the rows -1, 0, and +1), and $B_{fz}$ increases in the positive $x$-direction $\left(G = \partial B_{fz}/\partial x \big|_{x=0} > 0\right)$. The Larmor frequency of spin $p$ can be written as

$$\Omega_p \approx K\chi(x_p/a) + \Omega_{pd}, \quad \Omega_{pd} = \chi \sum_{k \neq p} A_{kp} m_{kz}. \qquad (8)$$

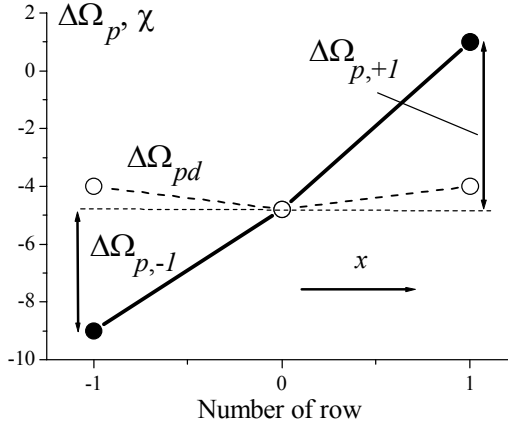
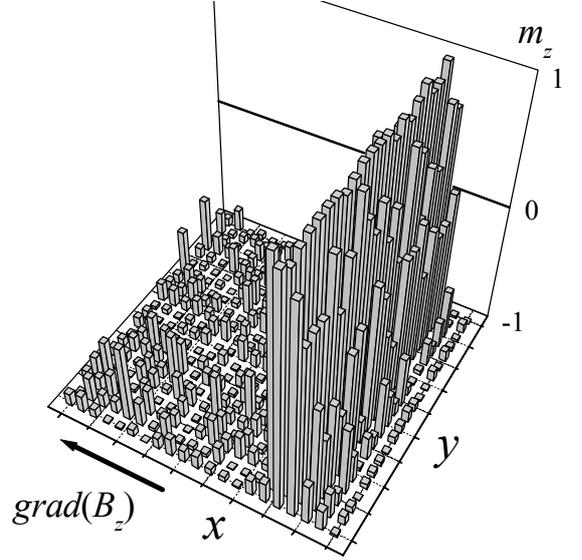

Fig. 5. Change of the Larmor frequencies in the neighboring rows −1, 0, 1. FRS is in the row "0". Circles show the change of frequencies due to the dipole-dipole interaction. Black circles show the frequency change due to the magnetic field gradient at $K = 5$. $\Delta\Omega_{p,-1} < \Delta\Omega_{p,+1}$.

Fig. 6. Distribution of $m_z$ at $\tau = 0.35$ for the relaxation process shown in Fig. 2, curve 2.

The initial value of $\Omega_{pd}$ (when all $m_{pz} \approx -1$) in the 2D-square spin lattice is $\Omega_{pd} \approx 8\chi$. Suppose that row "0" contains a single FRS. Initially the relaxation process occurs in row "0". The value of $m_{pz}$ in row "0" sharply changes to $m_{pz} \approx +1$. The corresponding changes of $\Omega_{pd}$ in rows 0, 1, and -1 are approximately the same:

$$\Delta\Omega_{pd,0} \approx -4.8\chi, \; \Delta\Omega_{pd,\pm 1} \approx -4\chi. \tag{9}$$

Spin diffusion is suppressed if the Larmor frequency difference due to magnetic field gradient becomes greater than $\Omega_{pd,0}$. This occurs roughly at $K > 4$. Our numerical simulations confirm this conclusion. (See Fig. 2, curve 2). At $K = 10$ the dependence $M_z(\tau)$ is approximately the same as for the system with no FRS. (See Fig. 2, "homogeneous" curve).

In a non-uniform magnetic field, spin relaxation is highly anisotropic. The relaxation process spreads first in the direction of the smaller magnetic field (because, for our simple example in Fig. 5, $\Delta\Omega_{p,-1} < \Delta\Omega_{p,+1}$) as occurred in 3D system [11]. This phenomenon can be used in a quantum computer. By changing the direction of the magnetic field gradient one can increase the relaxation time in any of the boundary chains (*N, S, W, E* in Fig. 1). As an example Fig. 6 demonstrates the distribution of $m_z$ at $\tau = 0.35$ for the relaxation process shown in Fig. 2, curve 2.

### 4. Spin relaxation in 1D- spin systems

In this section we study the relaxation process in 1D- spin systems interacting with a FRS. First, we consider a spin chain with $N = 41$ spins. The external magnetic field is perpendicular to the chain. We used two values $\chi = 1.3 \times 10^6$ and $\chi = 1.6 \times 10^5$, which correspond to the

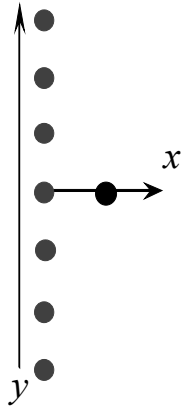

Fig. 7. FRS is placed at a distance $\sim a$ from the center of the chain.

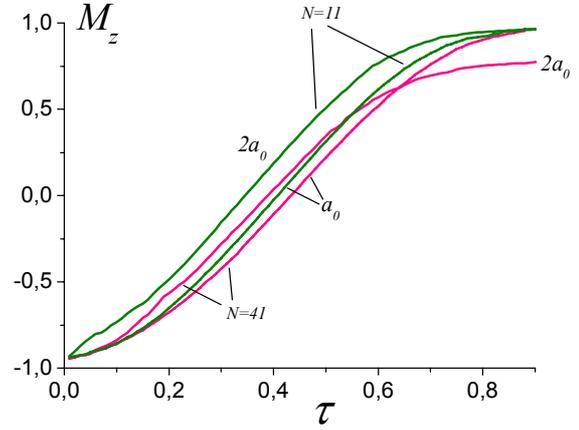

Fig. 8. Relaxation $M_z(\tau)$ in the spin chain for the number of spins in the chain $N = 11$ and $N = 41$. The distance between the spins equal to $a_0$ and $2a_0$

*a*

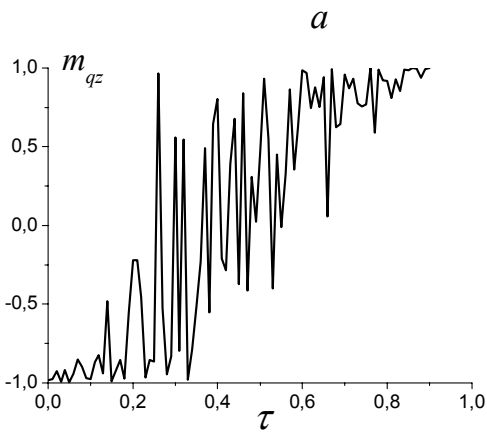

*b*

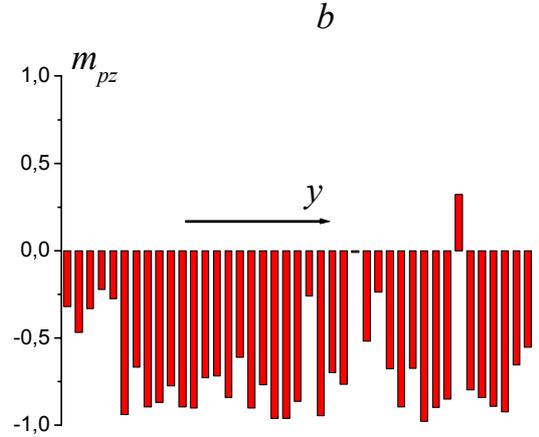

Fig. 9. *a* - Relaxation of the FRS $m_{qz}(\tau)$, *b* - distribution of $m_{pz}$ in the spin chain for $\tau = 0.2$. $N = 41$, $a = a_0 = 24 nm$.

distances $a = a_0 = 24 nm$ and $a = 2a_0 = 48 nm$. We put $\alpha = 0$, and $\beta/(N+1) = 5$. For these values of parameters in 3D- and 2D- spin system we calculate the relaxation time using Eq. (6) $\tau_1 \approx 2.2(N+1)/\beta = 0.44$.

If an FRS is at the distance of the order $a$ from the center of the chain (see Fig. 7) then the relaxation rate is similar to that in 3D- and 2D- systems. Fig. 8 demonstrates the relaxation $M_z(\tau)$ in the spin chain for the situation shown in Fig. 7.

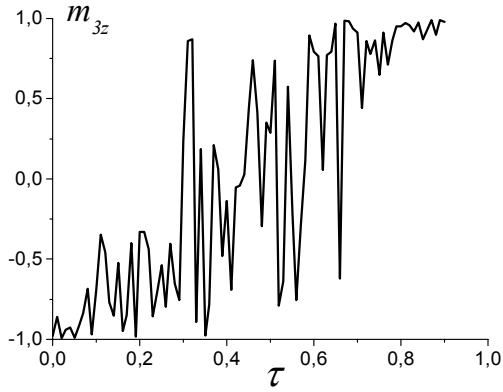

Fig. 10. Relaxation of the third spin in the spin chain. $N = 41$.

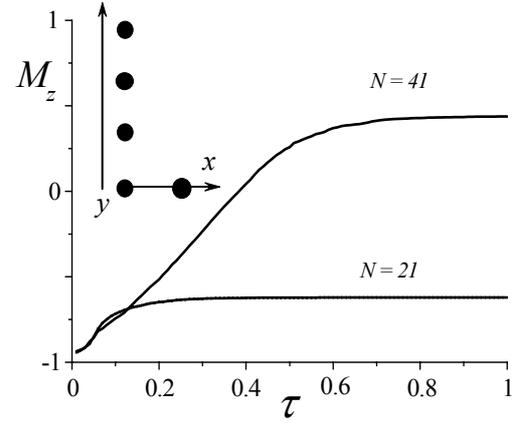

Fig. 11. Relaxation process $M_z(\tau)$ when FRS is placed near the end of the chain.

Fig. 9a demonstrates the dynamics of the FRS. (The value of the magnetic moment of the FRS $m_{qz}$ was taken in the time interval $\Delta \tau = 0.01$). Fig. 9b demonstrates the random distribution of the magnetic moments $m_{pz}$ in the spin chain at $\tau = 0.2$ ($M_z = -0.68$).

Note that the dynamics of a single spin in the chain is similar to the dynamics of an FRS. As an example, Fig. 10 shows relaxation of the third spin in the chain, $m_{3z}(\tau)$. If an FRS is placed near the end of the spin chain, the relaxation process in the spin chain slows down, and stops. (See Fig. 11). This "freezing" of the relaxation process is accompanied by the appearance of the stationary domain walls in the chain. (See Fig. 12 for $N = 21$ and Fig. 13 for $N = 41$). For $N = 21$ only those spins which are close to the FRS take part in the relaxation process. The relaxation process freezes with two domain walls for $N = 41$.

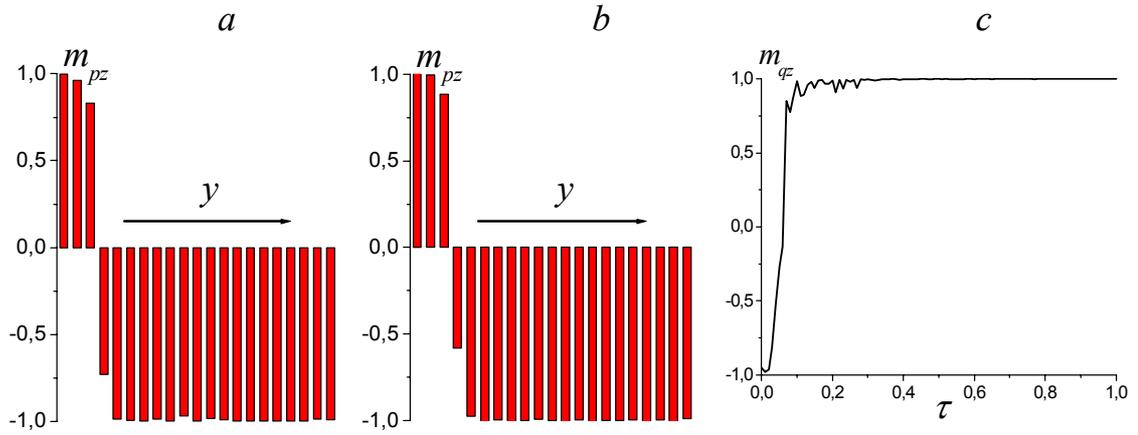

Fig. 12. *a* and *b* - Distribution of the magnetic moment $m_{pz}$ in the spin chain ($N=21$) at $\tau=0.3$ and $1$. *c* - Magnetic moment of FRS $m_{qz}$ as a function of $\tau$.

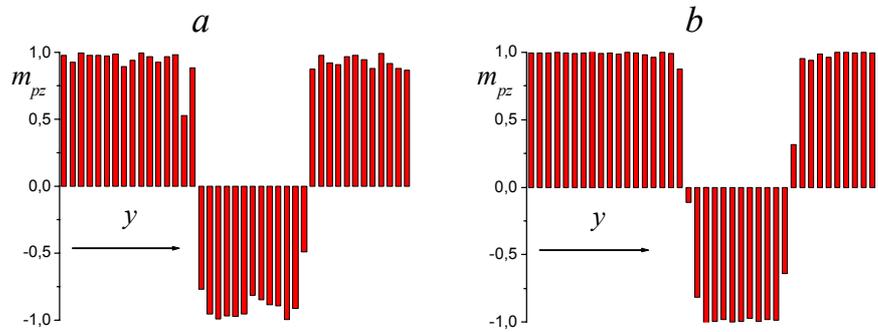

Fig. 13. Magnetic moment distribution at $\tau=0.6$ (*a*) and $\tau=1$ (*b*) for the spin chain with $N=41$.

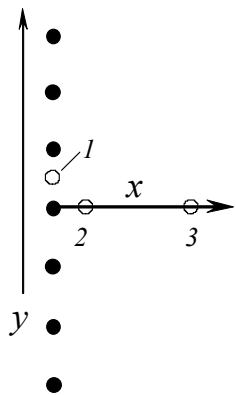 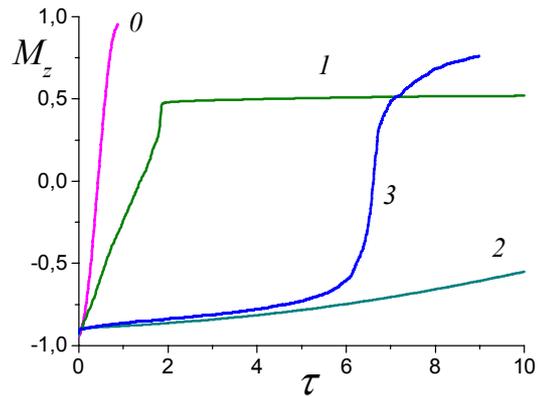

Fig. 14. FRS is very close to the spin chain (Position 1 and 2). In position 2, the distance from the chain $a_d = a/2$; in position 3, $a_d = 2a$.

Fig. 15. Relaxation of the magnetic moment of the chain $M_z(\tau)$ for the three positions of the FRS shown in Fig. 13. Curve "0" is taken from Fig. 7 for comparison. $N=21$.

Next, we consider the case in which the FRS and the chain spins which are close to FRS experience a higher dipole field than the other spins in the chain (Fig. 14, positions 1 and 2). Fig. 15 demonstrates the relaxation process for the three positions of the FRS shown in Fig. 14. One can see that the high inhomogeneity of the dipole field near the FRS suppresses the relaxation rate.

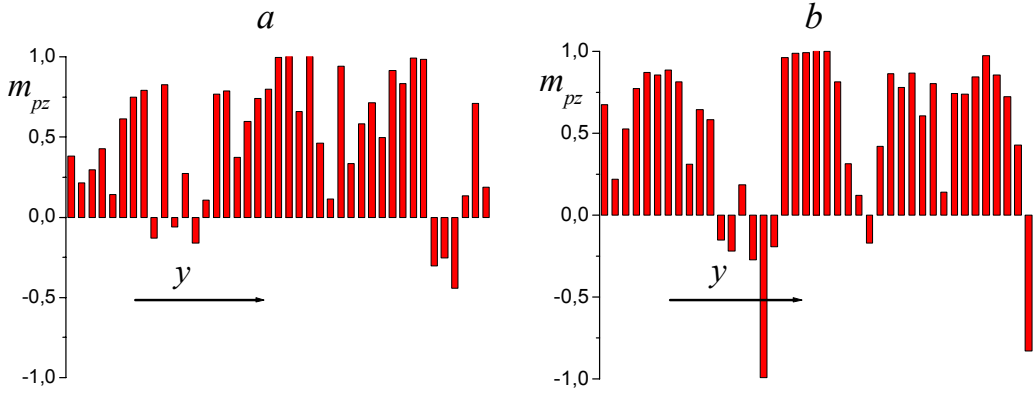

Fig. 16. Magnetic moment distributions corresponding to FRS position 1 in Fig. 13 for $\tau = 3$ (*a*) and $\tau = 7$ (*b*).

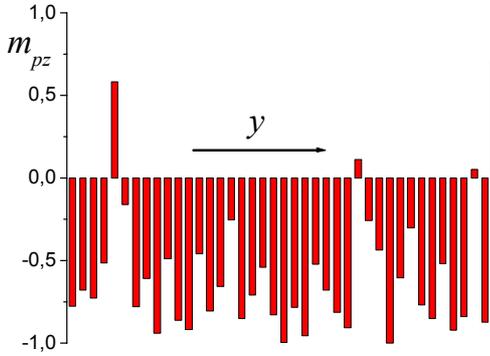

Fig. 17. Magnetic moment distributions corresponding to position 2 of FRS in Fig. 1 for $\tau = 10$.

Fig. 16 shows distributions of the magnetic moments corresponding to position 1 of FRS in Fig. 14 at two different time.

The distribution of the magnetic moments for a FRS in position 2 (see Fig. 14) is shown in Fig. 17. Note that in both Fig. 16 and 17 the distribution of the magnetic moments appears to be random rather than ordered.

For position 3 in Fig. 14 the dipole field on FRS is much smaller than the dipole field on the spins of the chain. The relaxation rate is suppressed initially but quickly increases at $\tau \approx 6$. (See curve 3 in Fig. 15). The characteristic feature of this case is the generation of a moving domain wall. (See Fig. 18). We have found that the direction of the motion of the domain wall depends on the position of FRS. If we place the FRS to the left of the chain in Fig. 14 then the domain wall will move in the negative *y*-direction. Thus, the process of spin relaxation in the spin chain appears to generate the random spin distribution, the quasi-stationary domain wall, and the moving domain wall.

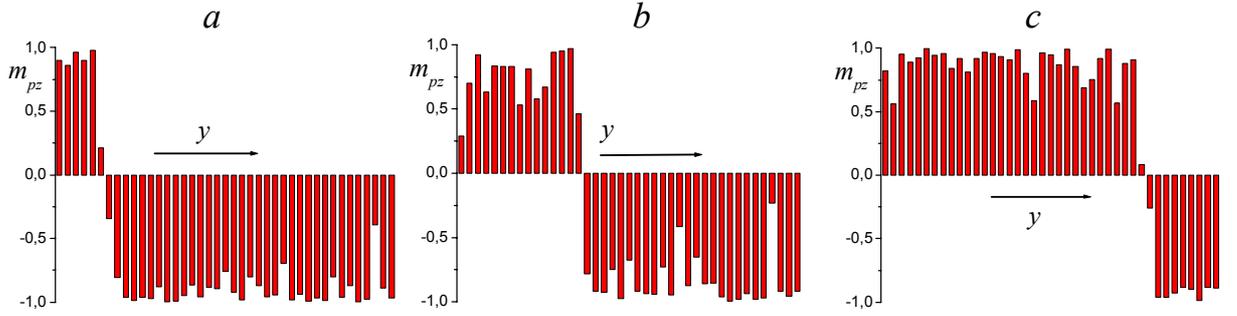

Fig. 18. "Moving domain wall" corresponding to position 3 of FRS in Fig. 13. $\tau = 6$ (*a*), 6.5 (*b*), 7 (*c*).

### 5. Suppression of spin diffusion and relaxation in 1D- spin systems

Finally, we consider the suppression of the spin diffusion and relaxation using a non-uniform magnetic field. We will discuss the situation corresponding to an FRS in location 2 in Fig. 14. Fig. 19 clarifies the spin dynamics in the uniform magnetic field for this situation.

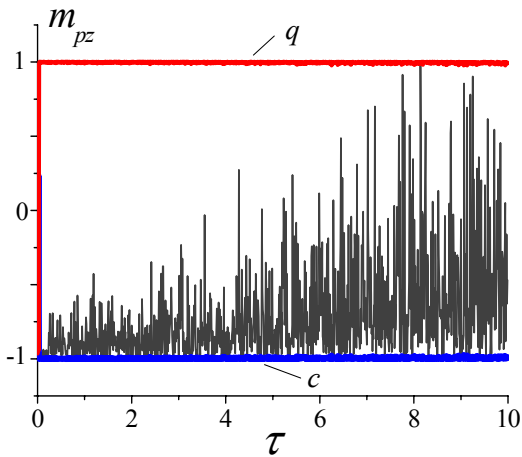 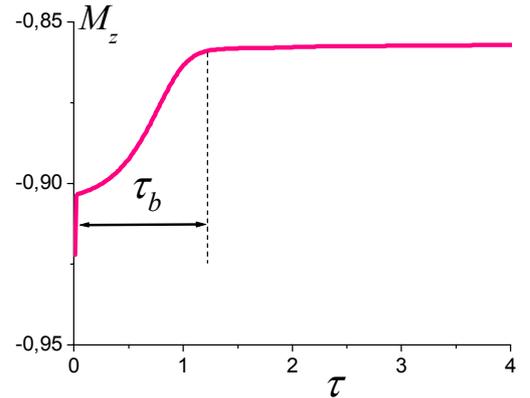

Fig. 19. Relaxation process for FRS in position "2" in Fig. 13. Horizontal lines "c" and "q" correspond to the central spin and FRS. Chaotic lines correspond to other spins of the chain. $N = 41$, $K = 0$.

Fig. 20. $M_z(\tau)$ for FRS in position "2" in Fig. 13. $N = 41$, $K = 50$. Magnetic field increases in the positive *x*-direction.

The central spin "c", which is close to the FRS quickly approaches the stable inverted state $m_{cz} = -1$ while FRS approaches to the stable ground state $m_{qz} = +1$ (horizontal lines "c" and "q" in Fig. 19).

For other spins, their magnetic moments $m_{pz}$ change chaotically. Fig. 20 demonstrates the relaxation of $M_z(\tau)$ in the spin chain in the presence of the non-uniform magnetic field. In this case the FRS magnetic moment $m_{qz}$ quickly approaches the value $m_{qz} = +1$, the central spin "c" also approaches the ground state $m_{cz} = +1$, while the other spins in the chain remain frozen. The change of $M_z(\tau)$ in Fig. 20 is caused by the relaxation of the central spin during the time interval $\tau_b$.

For a SQC, one should suppress the relaxation of the central spin. Relaxation of the FRS changes the dipole field on the neighboring spins of the chain to the value $-2\chi$. In order to suppress relaxation effectively the frequency difference between the spins in the chain must increase after FRS relaxation. Consequently, the magnetic field should increase in the positive *x*- direction in Fig. 14. Fig. 21*a* shows the relaxation of the central spin for three values of $K$, and Fig. 21*b* demonstrates the magnetic moment distribution for $K = 100$. If the distance $a_d$ between FRS and the central spin increases, then the FRS can be "isolated" at smaller values of $K$. In the previous figures we used $a_d = a/2$. Fig. 22 demonstrates the results of our simulations for $a_d = a$. One can see that for $K \geq 10$ the relaxation process is suppressed for every individual spin. Fig. 23 demonstrates the distribution of the magnetic moments for $K = 10$ and $K = 25$ at $\tau = 12$.

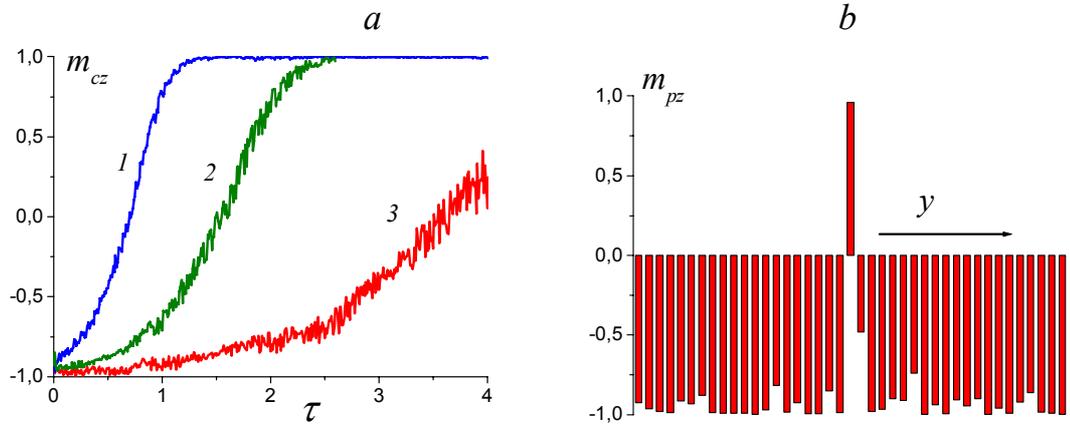

Fig. 21. *a*-Relaxation of the central spin for $K = 50$ (1), 70 (2), 100 (3). *b*-Distribution of magnetic moments at $\tau = 10$ for $K = 100$. (FRS is placed in the position "2" in Fig. 13, $N = 41$).

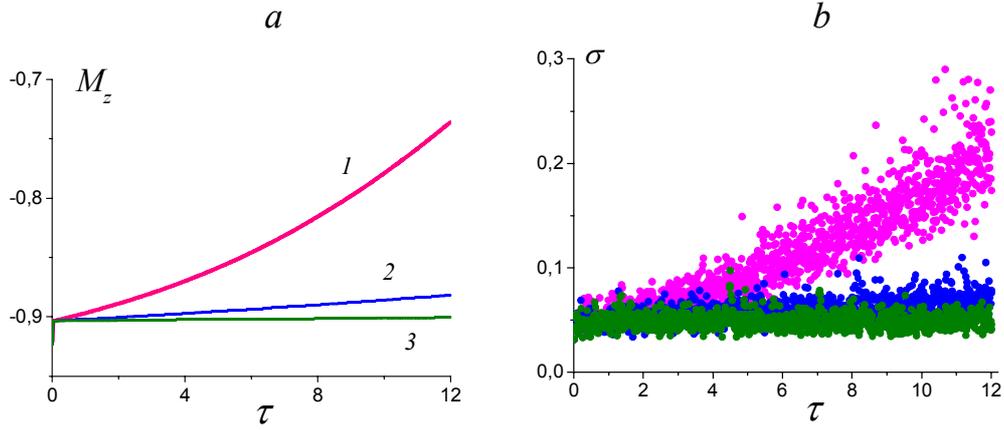

Fig. 22. $a$ – relaxation in the chain for $a_d = a$, $K = 5$ (curve 1), 10 (curve 2), 25 (curve 3). $b$ – mean square deviation $\delta(\tau) = \left[\dfrac{1}{N}\sum_p \left(m_{pz}(\tau) - M_z(\tau)\right)^2\right]^{1/2}$ as a function of $\tau$ for $K = 5$, 10 and 25.

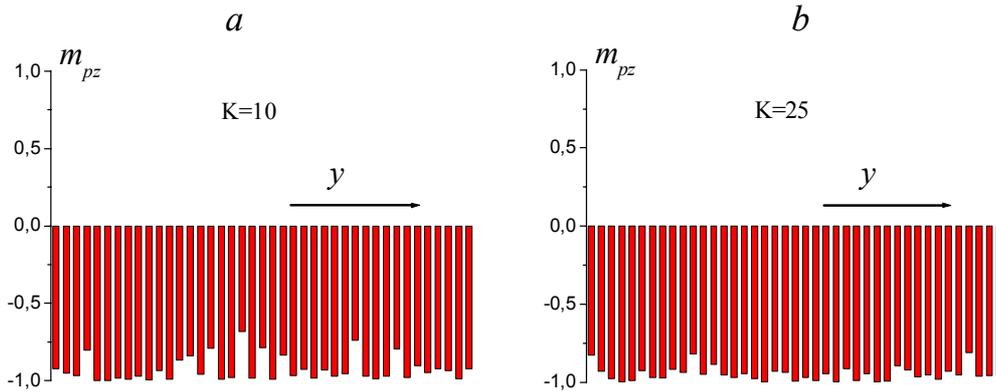

Fig. 23. Distribution of magnetic moments at $\tau = 12$ for $K = 10$ ($a$) and $K = 25$ ($b$).

Our semi-empirical estimate for the value of $K$ which is sufficient for the suppression of relaxation for the location of FRS along the $x$-axis (Fig. 14), can be written

$$K(a_d/a)^4 > 10. \tag{10}$$

Eq. (10) is satisfied for the gradient of the magnetic field $|\partial B_{fz}/\partial x| \sim 10^4 T/m$, and $a_d > 5nm$

## 5. Conclusion.

We have studied the processes of spin diffusion and relaxation for a SQC. We considered two possible implementations of SQC: (i) a boundary chain in a 2D- spin array, and (ii) an isolated spin chain. In both cases the FRS can appear outside the SQC spin chain. This causes relaxation in the SQC due to spin diffusion generated by the dipole-dipole interaction between the spins. We have demonstrated that in the first case, when the number of spins $N$ does not exceed a threshold value, the processes of spin diffusion and relaxation are similar to the case of a 3D- system, and the influence of the FRS does not depend on its location. In the second case, we have calculated various scenarios of spin diffusion and relaxation that include random distribution of the magnetic moments in SQC as well as formation of the stationary or moving domain walls. We have shown that the relaxation process can be suppressed by a non-uniform external magnetic field, and we have presented the characteristic estimate of parameters for this suppression.

## Acknowledgement

We thank G.D. Doolen for useful discussions. This work was supported by the Department of Energy (DOE) under Contract No. W-7405-ENG-36, by the National Security Agency (NSA), and by the Advanced Research and Development Activity (ARDA)